\documentclass{revtex4}
\usepackage{amsmath}
\usepackage{amssymb}
\usepackage{hyperref}
\usepackage{graphicx}
\usepackage{wasysym}

\begin{document}

\title{Cosmological scenarios in quantum cosmology of Brans-Dicke theory of gravity on Einstein and Jordan frames} 

\vspace{0.3cm}
\author{C.R. Almeida$^{(a)}$}\email{carlagbjj@hotmail.com} 
\author{A.B. Batista$^{(a)}$}\email{abrasilb918@gmail.com} 
\author{J.C. Fabris$^{(a,b)}$}\email{fabris@pq.cnpq.br}
\author{N. Pinto-Neto$^{(c)}$}\email{nelson.pinto@cnpq.br}
\vspace{0.5cm}

\bigskip

\affiliation{$^{(a)}$ Universidade Federal do Esp\'{\i}rito Santo, CEP 29075-910, Vit\'{o}ria/ES, Brazil}

\affiliation{$^{(b)}$ National Research Nuclear University “MEPhI”, Kashirskoe sh. 31, Moscow 115409, Russia }

\affiliation{$^{(c)}$ Centro Brasileiro de Pesquisas F\'{\i}sicas, Rio de Janeiro CEP22290-180, RJ, Brazil}

\affiliation{...}
\affiliation{...}




\begin{abstract}
The Hamiltonian of any quantum cosmological model must obey some conditions in order to be self-adjoint, or to admit self-adjoint extensions. We will consider in this paper the self-adjoint character of the Hamiltonian resulting from the Brans-Dicke theory, both in Einstein and Jordan frames. In both cases we find the wave-function solutions in order to obtain concrete predictions for the evolution of the Universe. We show that the specific cosmological scenarios obtained depend very weakly on the conditions of having a self-adjoint operator. The problem of equivalence between Einstein and Jordan frames is considered, and it is shown that this equivalence implies a specific ordering parameter, as well as a particular choice of the physical variables.
\end{abstract}

\maketitle

\section{Introduction}

In ordinary quantum mechanics it is necessary that an operator must be self-adjoint in order to represent a physical observable. This necessity is connected with the usual interpretation framework used in quantum mechanics, i.e., the Copenhagen Interpretation. Other interpretations of quantum mechanics, like the ontological interpretation, may not require the operators to be self-adjoint, even though this notion remains one of the most important in the mathematical formulation of a quantum theory \cite{neu}. It must be stressed that a self-adjoint operator is more than simply a Hermitian (or symmetric) operator, since the self-adjointness concept requires, besides that an operator must be symmetric, that its domain must be the same as that of its Hermitian conjugate. For a clear mathematical exposition of these concepts, see Ref. \cite{Reed&Simon}. For a discussion of the different interpretation schemes of quantum mechanics, see Ref. \cite{holland,nelson}.

Quantum cosmology is the quantization of the Universe as a whole. Such ambitious program faces naturally many conceptual and technical challenges. For some review on this topic, see Refs. \cite{halliwell,nelson-fabris,bojo}. An important issue is the absence of a clear time coordinate. This is due to the fact that, in the ADM Hamiltoninian formulation of General Relativity, the invariance with respect to time reparametrization imply the constraints,
\begin{eqnarray}
{\cal H} \approx 0,
\end{eqnarray}
which leads, after quantization, to a time-independent equation. This problem can be circumvented either by identifying an ''internal'' time parameter (what is generally called deparametrization) or by introducing matter fields that can play the r\^ole of time. Another important issue in quantum cosmology
is the problem of the collapse of the wavefunction. This touches the foundations of quantum mechanics, and requires either a reformulation of the usual Copenhagen interpretation or the use of a complete different framework as the one proposed in the ontological interpretation. The many-worlds interpretation of quantum mechanics \cite{m-w} is another possible reformulation of the Copenhagen interpretation, also adapted to the quantum cosmology context.

In some simple configurations, the self-adjoint character of the Hamiltonian operator resulting from quantum cosmology in mini-superspace can be easily determined. In Refs. \cite{gotay,rubakov,lemos1, lemos}, the case where gravitation is minimally coupled to matter and using Schutz formalism \cite{schutz1,schutz2} to describe the fluid in order to obtain a time-variable, the self-adjoint properties of the effective Hamiltonian has been analyzed. Boundary conditions must be imposed in order to have a self-adjoint Hamiltonian, as it happens with the problem of a particle defined in the half real line \cite{Reed&Simon}.

The case of scalar-tensor theories of gravity are much more involved concerning their application to quantum cosmology. These theories contain a non-minimal coupling between a scalar field an the gravitational sector. This non-minimal coupling introduces a more involved set of equations. In particular, after quantization in mini-superspace, besides having now two variables (the scale factor and the scalar field), it appears cross-derivatives terms with respect to these two variables which are directly related to the non-minimal coupling. These technical aspects can be circumvented in many different ways. One of them is to perform a conformal transformation to a frame where the non-minimal coupling (Jordan frame) gives place to a minimal coupling between gravity and the scalar sector (Einstein frame). In general, the conformal transformation simplifies the structure of the gravitational sector, but at the price of introducing a direct coupling of the scalar field with the matter sector.
This problem does not exist if the matter sector contains only a radiative fluid, which is conformally invariant.

The prototype of a scalar-tensor theory is the Brans-Dicke theory \cite{bd}. It is also the simplest one. In reference \cite{moniz2}, quantization of
the Brans-Dicke theory in mini-superspace has been analyzed, introducing also a radiative fluid described through the Schutz formalism, which plays the r\^ole of a time variable. The analysis has been done in the Einstein frame. Ordering factors have been introduced in the resulting Schr\"odinger-like equation to take into account the ambiguity in the order of the momentum and position operators during the quantization procedure. In that frame, the self-adjointness of the effect Hamiltonian has been investigated. Considering the physical domain of square-integrable wave functions, for some range of the order parameters the Hamiltonian operator is self-adjoint, while for other ranges it is not self-adjoint but it admits self-adjoint extensions. 
There are cases where the Hamiltonian is neither self-adjoint, nor admits a self-adjoint extension.

Quantum cosmological models derived from the Brans-Dicke theory have been treated in many different contexts, including its reformulation in a minimally coupled frame \cite{fnc1,fnc2}. In the study carried out in Ref. \cite{moniz2}, the determination of specific cosmological scenarios has not been addressed (this problem has been partially addressed in Ref. \cite{moniz}). Moreover, the properties of the effective Hamiltonian have been studied in details only in the Einstein frame. The present work will try to fill these gaps, and to complete and generalize those previous studies. First, we will revise the results obtained in \cite{moniz2,moniz} (sections 2 and 3), and subsequently the cosmological scenarios emerging from the quantum model, in the Einstein frame, will be studied (section 4). It will be shown that bouncing models are predicted, whose forms depend slightly on the ordering parameter, hence on the self-adjoint character of the effective Hamiltonian  (essentially, some constants change their value according to the ordering parameters). In section 5, we analyze the same problems in the Jordan frame.
It comes out that only for some very specific values of the ordering parameters the operator is self-adjoint. These values correspond to the situation where there is
the equivalence between the Einstein and Jordan frame. Moreover, our results suggest that the variables defined in Einstein frame are the good observables of the theory. This seems to point out that the Einstein frame is the physical frame of such quantum versions of the Brans-Dicke theory. The possible limitations of this analysis, as well as its possible extensions, are discussed in the Conclusions (section 6).

\section{The model}

The Brans-Dicke Lagrangian, with a non-minimally coupled scalar field is given by
\begin{equation}
\label{BD Lagrangian}
\mathcal{L}_{G} = \sqrt{- \tilde g} \left(\phi \tilde R - \omega \frac{\phi_{;\rho} \phi^{\rho}}{\phi}\right) \,,
\end{equation}
where $\omega$ is the Brans-Dicke constant. From a conformal transformation, $\tilde{g}_{\mu \nu} = \phi^{-1} \, g_{\mu \nu}$, where $\tilde{g}_{\mu \nu}$ is the metric in the non-minimal coupling frame, the Lagrangian reads as
\begin{equation}
\label{GR Lagrangian}
\mathcal{L}_{G} = \sqrt{-g} \left[R - \biggr(\omega + \frac{3}{2}\biggl) \frac{\phi_{;\rho} \, \phi^{;\rho}}{\phi^{2}}\right] \,,
\end{equation}
which is the Lagrangian for General Relativity with a scalar field minimally coupled. The Lagrangian \eqref{BD Lagrangian} is written in the Jordan frame, and \eqref{GR Lagrangian} is written in the Einstein frame.

The Lagrangian (\ref{GR Lagrangian}) can be rewritten as
\begin{eqnarray}
\mathcal{L}_{G} = \sqrt{-g} \left[R - \epsilon\biggr|\omega + \frac{3}{2}\biggl| \sigma_{;\rho} \, \sigma^{;\rho}\right] \,,
\end{eqnarray}
where
\begin{eqnarray}
\phi = e^\sigma \quad \text{and} \quad \epsilon = \mbox{sig} \biggr(\omega + \frac{3}{2}\biggl) \,.
\end{eqnarray}
This Lagrangian can also be written in the Einstein frame as,
\begin{eqnarray}
\mathcal{L}_{G} = \sqrt{-g} \left[R - \epsilon \bar\sigma_{;\rho} \, \bar\sigma^{;\rho}\right] \,,
\end{eqnarray}
with
\begin{eqnarray}
\bar\sigma = \sqrt{\biggr|\omega + \frac{3}{2}\biggl|}\sigma.
\end{eqnarray}

In the Jordan frame, the Friedmann-Lema\^{\i}tre-Robetson-Walker (FLRW) flat metric describing an isotropic and homogeneous universe is given by,
\begin{eqnarray}
ds_J^2 = dt^2 - a(t)^2(dx^2 + dy^2 + dz^2),
\end{eqnarray}
where $a(t)$ is the scale factor. The FLRW metric in the Einstein frame is given by,
\begin{eqnarray}
ds_E^2 = d\bar t^2 - b(\bar t)^2(dx^2 + dy^2 + dz^2),
\end{eqnarray}
where $b(\bar t)$ is the scale factor in this frame. Writing the scalar field as,
\begin{eqnarray}
\label{tc1}
\phi = e^{\bar\sigma},
\end{eqnarray}
the two scale factors and the two time coordinates are related by,
\begin{eqnarray}
\label{tc2}
a = e^{-\frac{\bar\sigma}{2}}b, \quad \bar{t} = e^{-\frac{\bar\sigma}{2}}t.
\end{eqnarray}

We introduce now a radiative perfect fluid as a matter component. Using the Schutz's variable, it is possible to show that the momentum associated to the the radiative fluid appears linearly in the Hamiltonian. This procedure is described in details in reference \cite{lemos}. Hence, after quantization, the fluid variable plays the r\^ole of time. Quantizing the system in the Jordan frame, the Schr\"odinger-type equation reads,
\begin{eqnarray}
\label{ej}
\partial_a^2\Psi + \frac{p}{a}\partial_a\Psi + \frac{6}{\omega}\biggr[\frac{\phi}{a}\partial_\phi\partial_a\Psi - \frac{1}{a^2}\biggr(\phi^2\partial_\phi^2\Psi + q\phi\partial_\phi\Psi\biggl)\biggl] =
12i\frac{(3 + 2\omega)}{\omega}\phi\partial_T\Psi,
\end{eqnarray}
where $T$ is the time variable emerging, after quantization, from the application of the Schutz formalism to the radiative fluid, while $p$ and $q$ are ordering factors appearing in the quantization process due to the non-commutativity of the position and momentum operators. Cross derivatives with respect to $a$ and $\phi$ appear due the non-minimal coupling. For the moment, $p$ and $q$ are arbitrary.

The quantization of this model in the Einstein frame was done in \cite{moniz2} and the resulting Scrh\"odinger equation is given by
\begin{equation}
\label{Schroedinger equation}
- \partial_{b}^{2} \Psi - \frac{\bar p}{b} \partial_{b} \Psi + \frac{\varpi}{b^{2}} \left( \phi^{2} \partial_{\phi}^{2} \Psi + \bar q\phi \partial_{\phi} \Psi \right) = i \partial_{T} \Psi \,,
\end{equation}
The constants $\bar p$ and $\bar q$ are also ordering factors. They are related to the ordering factors in the Jordan frame by,
\begin{eqnarray}
\label{relation p overlinep}
 p = \frac{9 + 2\omega \bar p -6 \bar q }{3 + 2\omega}, \quad q = \bar q.
\end{eqnarray}
The constant $\varpi$ is related to the Brans-Dicke constant $\omega$ in the following way
\begin{equation}
\varpi = \frac{12}{3+2\omega} \,.
\end{equation}
Note that, in the Einstein frame, the Schr\"odinger equation has no cross derivative term, as we would expect.
The choice of the radiative matter is due to the fact that it is conformaly invariant, thus the time parameter does not change in both frames.

For now, let us firstly find the solutions of eq. \eqref{Schroedinger equation} and find the expected values for the scale factor and scalar field. The corresponding solutions in the Jordan frame can than be determined. Afterward, we will investigate the quantum equivalence between Einstein and Jordan frames.

\section{The solutions}

The Schr\"odinger equation \eqref{Schroedinger equation} has solutions $\Psi(b,\phi,T) = \psi(b,\phi) \, e^{iET}$, where the function $\psi$ satisfies the eigenvalue equation,
\begin{eqnarray}
\label{eigenvalue eq.}
\biggr\{\partial_b^2 + \frac{\bar p}{b}\partial_ b - \varpi \frac{\phi^2}{b^2}\biggr[\partial_\phi^2 + \frac{1}{\phi}\partial_\phi\biggl]\biggl\}\psi = - E\psi \,.
\end{eqnarray}
We have fixed $q = \bar q = 1$ since, as explained in reference \cite{moniz2}, it is the only case were we can have square integrable functions in the variable $\phi$.

Let us define $\epsilon$ and $\sigma$ as
\begin{eqnarray}
\label{definition epsilon and sigma}
\varpi = \epsilon|\varpi| \quad;\quad \sigma = \frac{\ln\phi}{\sqrt{|\varpi|}} \,.
\end{eqnarray}
Thus, the equation becomes,
\begin{eqnarray}
\label{eigenvalue eq. sigma}
\biggr\{\partial_b^2 + \frac{\bar p}{b}\partial_ b - \epsilon \frac{1}{b^2}\partial_\sigma^2\biggl\}\psi = - E\psi \,.
\end{eqnarray}
Using the separation of variable method, the wave function is defined as $\psi (b, \sigma) = X(b)Y(\sigma)$. Then, the partial differential equation \eqref{eigenvalue eq. sigma} becomes a sistem of ordinary differential equations:
\begin{eqnarray}
\label{se1}
\biggr\{\partial_b^2 + \frac{\bar p}{b}\partial_b + E + \epsilon \frac{k^2}{b^2}\biggl\}X &=& 0 \,,
\\
\label{se2}
Y'' + k^2 Y &=& 0 \,.
\end{eqnarray}
The primes mean derivatives with respect to $\sigma$. We must fix the sign of the separation constant as $- k^2$ because otherwise there is no square integrable solution (see \cite{moniz2}).

The solutions of equations \eqref{se1} and \eqref{se2} are, respectively:
\begin{eqnarray}
\label{solution X}
X(b) &=& b^{\frac{1}{2}(1 - \bar p)}\biggr\{A_1 J_\nu(\sqrt{E}b) + A_2 J_{-\nu}(\sqrt{E}b)\biggl\} \,,
\\
\label{solution Y}
Y(\sigma) &=& B_1 e^{ik\sigma} + B_2 e^{-ik\sigma} \,,
\end{eqnarray}
where $A_{1,2}$ and $B_{1,2}$ are integration constants. Moreover,
\begin{eqnarray}
\label{definition nu}
\nu = \sqrt{\biggr(\frac{\bar p - 1}{2}\biggl)^2 - \epsilon \, k^2} \,.
\end{eqnarray}
The resulting wavefunction is:
\begin{eqnarray}
\label{general solution}
\Psi(b,\sigma,T) = b^{\frac{(1 - \bar p)}{2}}\biggr\{A_1 J_\nu(\sqrt{E}b) + A_2 J_{-\nu}(\sqrt{E}b)\biggl\}\left( B_1 e^{ik\sigma} + B_2 e^{-ik\sigma} \right)e^{iET}\,.
\end{eqnarray}
To determine the constants $A_{1,2}$ and $B_{1,2}$, we must impose boundary conditions on the wave solutions.

\section{The boundary conditions and the expectation value in the Einstein frame}
\label{section EF}

In the Copenhagen Interpretation of Quantum Mechanics, the operator that describes observables must be self-adjoint in order to produce measurements with real outcomes (eigenvalues). For an operator to be self-adjoint, it has to be symmetric and its domain should coincide with the domain of its adjoint, see reference \cite{Reed&Simon} for a clear exposition of the properties of self-adjoint operators. In \cite{moniz2}, there is a complete description of the problem of self-adjointness in the canonical quantization of the Brans-Dicke theory, on which it is shown that the Hamiltonian operator
\begin{eqnarray}
\label{Hamiltonian operator}
\hat{H} = - \partial^2_b - \frac{\bar p}{b}\partial_b + \epsilon \frac{1}{b^2} \partial_\sigma^2
\end{eqnarray}
is symmetric if the internal product is defined as,
\begin{eqnarray}
\label{internal product}
(\phi,\psi) = \int_0^\infty \int_{-\infty}^{+\infty}\phi^*\psi \,b^{\bar p} \, dbd\sigma \,.
\end{eqnarray}
Note that the measure $d\mu = b^{\bar p} \, dbd\sigma$ depends on the ordering factor $\bar p$.

In order for the operator \eqref{Hamiltonian operator} to be Hermitian, i.e. symmetric, we must have,
\begin{eqnarray}
\label{symmetric}
(\Psi, \hat{H}\Psi) = (\hat{H} \Psi,\Psi) \,.
\end{eqnarray}
Applying the previous expressions for the Hamiltonian and for the inner product, we find that the functions $\Psi$ in the domain of $\hat{H}$ must satisfy the boundary conditions
\begin{eqnarray}
\label{boundary conditions1}
\biggr\{\int_{0}^{+\infty}\biggr[\Psi^*\partial_\sigma\Psi - \partial_\sigma\Psi^*\Psi\biggl]b^{\bar p} db\biggl\}_{\sigma=-\infty}^{\sigma=+\infty} &=& 0 \,;
\\
\label{boundary conditions2}
\biggr\{\int_{-\infty}^{+\infty}\biggr[\Psi^*\partial_b\Psi - \partial_b\Psi^*\Psi\biggl]b^{\bar p} d\sigma\biggl\}_{b=0}^{b=+\infty} &=& 0
\end{eqnarray}
in order that Eq.~\eqref{symmetric} be verified. The main point now is, taking into account the solution for the wavefunction, to verify the conditions in which the boundary terms are equal to zero.

Solution \eqref{general solution} generally does not satisfy none of the boundary conditions \eqref{boundary conditions1} and \eqref{boundary conditions2}. However, considering $A_2 = B_2 = 0$ to construct a wave packet with the form
\begin{eqnarray}
\Psi(b,\sigma,T) = \int_{0}^{\infty} \int_{-\infty}^{\infty} b^{\frac{1}{2}(1 - \bar p)}A(k,E)\,J_\nu(\sqrt{E}b)\,e^{ik\sigma}\,e^{iEt} \, dkdE \,,
\end{eqnarray}
with $A(k,E)$ being the normalization function, the conditions \eqref{boundary conditions1} and \eqref{boundary conditions2} are satisfied. By choosing $A(k,E)$, as a gaussian function and considering $E=x^{2}$, the
wave packet becomes,
\begin{eqnarray}
\Psi(b,\sigma,T) = b^{\frac{1}{2}(1 - \bar p)}  \int_{0}^{\infty} \int_{-\infty}^{+\infty} e^{-k^2} x^{\nu + 1} e^{- \alpha x^2}J_\nu(xb)\, e^{ik\sigma} \, dkdx \,,
\end{eqnarray}
where $\alpha = \gamma - iT$, with $\text{Re}(\gamma) >0$. After integration in $x$ \cite{Gredshteyn}, the wave packet reads,
\begin{eqnarray}
\label{wavepacket}
\Psi(b,\sigma,T) = b^{\frac{1}{2}(1 - \bar p)} \int_{-\infty}^{+\infty} e^{- k^2}e^{ik\sigma} \, \frac{b^\nu}{(2\alpha)^{\nu + 1}} \,e^{- \frac{b^2}{4\alpha}} \,dk \,.
\end{eqnarray}

Using the internal product \eqref{internal product}, the norm of this wave packet \eqref{wavepacket} reads
\begin{eqnarray}
N &=& \int_{-\infty}^{+\infty} \int_{-\infty}^{+\infty} \int_{-\infty}^{+\infty} \int_0^{+\infty} e^{- k^2 - {k'}^2} e^{i(k-k^{\prime}) \sigma} \nonumber
\\
&\times&\frac{b^{\nu +{\nu'}^* + 1}}{(2\alpha)^{\nu + 1}(2\alpha^*)^{{\nu'}^* + 1}} \, e^{- \gamma\frac{b^2}{4\alpha\alpha^*}} \, db\, d\sigma \, dk\,dk' \,.
\end{eqnarray}
The integration over $\sigma$ generates a delta function and, after integrating over $k^{\prime}$, we obtain:
\begin{eqnarray}
\label{norm}
N = \int_{-\infty}^{+\infty}\int_0^{+\infty}e^{- 2k^2}\frac{b^{\nu +{\nu}^* + 1}}{(2\alpha)^{\nu + 1}(2\alpha^*)^{{\nu}^* + 1}} \,e^{- \gamma\frac{bˆ2}{4\alpha\alpha^*}} \, db\,dk \,.
\end{eqnarray}
For the moment, let us continue with an arbitrary ordering factor $\bar p$.

Let us concentrate in the case $\epsilon = -1$. In this case, the Hamiltonian operator is bounded from below, and it is essentially self-adjoint for $-1 \leq \bar p \leq 3$ (see \cite{moniz2}). For $\epsilon = 1$, there is an ambiguity in the range of the variables, due to the fact that the energy $E$ is not bounded from below. For this same reason, it is possible that the configurations are unstable. We will not consider the case $\epsilon = 1$ in the rest of this work.

If $\epsilon = - 1$, $\nu \in \mathbb{R}$. The above expression \eqref{norm} reduces to,
\begin{eqnarray}
N = \int_{-\infty}^{+\infty}\int_0^{+\infty}e^{- 2k^2}\frac{b^{2\nu + 1}}{(4\alpha\alpha^*)^{\nu + 1}} \, e^{- \gamma \frac{b^2}{4\alpha\alpha^*}} \, db\, dk\,.
\end{eqnarray}
Defining new variables as,
\begin{eqnarray}
\label{variable y}
y = \frac{b}{\sqrt{B}} \quad; \quad B = \alpha\alpha^* = \gamma^2 + T^2 \,,
\end{eqnarray}
the norm becomes,
\begin{eqnarray}
N = \int_{-\infty}^{+\infty}\int_0^{+\infty}e^{- 2k^2} \frac{y^{2\nu + 1}}{4^{\nu + 1}}e^{- \gamma \left(\frac{y}{2} \right)^{2}} \, dy \, dk\,,
\end{eqnarray}
which is time-independent.

We can now compute the expectation value for $b$ and $\sigma$. For $b$ we have, following the same steps as before,
\begin{eqnarray}
\langle b \rangle &=& \frac{1}{N}\int_{-\infty}^{+\infty} \int_{0}^{\infty}  e^{-2k^2} \, \frac{y^{2\nu + 1}}{4^{\nu + 1}} \, b\, e^{-\frac{\gamma}{4} y^{2}} \, dy \,dk
\nonumber
\\
&=& \frac{\sqrt{\gamma^2 + T^2}}{N} \int_{-\infty}^{+\infty} \int_{0}^{\infty}  e^{-2k^2} \,\frac{y^{2\nu + 2}}{4^{\nu + 1}} \,y \,e^{-\frac{\gamma}{4}y^{2}} \, dy \, dk\,.
\label{expected value b}
\end{eqnarray}
This means that the scale factor behaves as,
\begin{eqnarray}
\langle b \rangle \propto \sqrt{\gamma^2 + T^2} \,,
\end{eqnarray}
in agreement with \cite{moniz}. Note that, since $\gamma^{2} > 0$, the scale factor $b$ is always greater than zero, the volume of the universe never shrinks to zero. Hence, the model is free of singularities: the universe contracts, bounces at a minimal scale factor, and
expands afterwards. This result is valid for any value of the ordering parameter $\bar p$.

The computation of the expectation value of the scalar field \footnote{Remeber that $\sigma$ is not actually the scalar field, but its logarithm.} is more involved. Now, we have,
\begin{eqnarray}
\langle \sigma \rangle &=& \frac{1}{N} \int_{-\infty}^{+\infty} \int_{-\infty}^{+\infty} \int_{-\infty}^{+\infty} \int_{0}^{\infty} \, e^{-k^2 - {k'}^2} \sigma \, e^{i(k - k')\sigma}
\nonumber
\\
&\times& \frac{b^{\nu + \nu' + 1}}{(2\alpha)^{\nu + 1}(2\alpha^*)^{\nu'+ 1}} \, e^{- \frac{\gamma}{4}\frac{b^2}{\alpha\alpha^*}} \, db \, d\sigma \, dk\,dk'\,.
\end{eqnarray}
Considering the identity
\begin{equation}
\sigma \, e^{i(k - k')\sigma} = -i \partial_{k} \, e^{i(k - k')\sigma} \,,
\end{equation}
we integrate over $\sigma$, to obtain a delta function. Using the property $\delta^{\prime} \times f = \delta \times f^{\prime}$, we obtain:
\begin{eqnarray}
\langle \sigma \rangle = -\frac{i}{N} \int_{-\infty}^{+\infty} \int_{0}^{\infty} e^{- \frac{\gamma}{4} \frac{b^2}{\alpha\alpha^*}} \, e^{-2k^2}\frac{b^{\nu  + 1}}{(2\alpha)^{\nu + 1}(2\alpha^*)}
\biggr[\frac{d\nu}{dk}\ln\biggr(\frac{b}{2\alpha^*}\biggl) - 2k\biggl]db \,dk\,.
\end{eqnarray}
Remembering that, for this case, $\nu = \sqrt{[(\bar p - 1)/2]^{2} + k^{2}}$, the integral clearly is zero for $\bar p \neq 1$. For $\bar p = 1$, $\nu = |k|$, and we find,
\begin{eqnarray}
\langle \sigma \rangle = - \frac{i}{N} \int_{-\infty}^{+\infty} \int_{0}^{\infty} \frac{e^{-2k^2}}{4^{|k| + 1}}y^{2|k| + 1} \, e^{- \frac{\gamma}{4} y^2} \frac{k}{|k|} \ln\biggr[\frac{y}{4} \biggr(\frac{\alpha}{\alpha^*} \biggl)^\frac{1}{2} \biggl] dy \, dk \,,
\end{eqnarray}
which is also zero. Hence, the expectation value of $\sigma$ is zero. This results stands even for $\bar p \neq 1$. This does not mean, of course, that $\sigma$ is zero in all realizations, since it is an expectation value, and fluctuations are allowed.

However, the null result for the $<\sigma>$ depends directly on the form of the superposition factor $A(k)$, in special to the fact we have chosen an even function.
If, instead, we had chosen a superposition factor with other parity, the result would be different. For example, choosing,
\begin{eqnarray}
A(k) = e^{- k - k^2},
\end{eqnarray}
we find the expression
\begin{eqnarray}
\langle \sigma \rangle = - \frac{i}{N} \int_{-\infty}^{+\infty} \int_{0}^{\infty} \frac{e^{- 2k - 2k^2}}{4^{|k| + 1}}y^{2|k| + 1} \, e^{- \frac{\gamma}{4} y^2} \frac{k}{|k|} \ln\biggr[\frac{y}{4} \biggr(\frac{\alpha}{\alpha^*} \biggl)^\frac{1}{2} \biggl] dy \, dk \,.
\end{eqnarray}
Now, the integrand has no definite parity. Writing,
\begin{eqnarray}
\alpha = \gamma + i T = Re^{i\theta}, \quad R = \sqrt{\gamma^2 + T^2}, \quad \theta = \arctan\biggr(\frac{T}{\gamma}\biggl),
\end{eqnarray}
the expectation value for $\sigma$ reads,
\begin{eqnarray}
<\sigma> = \sigma_0 + \sigma_1\arctan\biggr(\frac{T}{\gamma}\biggl),
\end{eqnarray}
where $\sigma_0$ and $\sigma_1$ are constants. Remarkably, these changes in the superposition factor does not alter the behaviour for the expectation value of the scale factor in the Einstein frame, $b$.

\section{Expectation values in the Jordan frame}

Let us return back to the Schr\"odinger equation in the Jordan frame. We will discuss about the conditions which must be imposed to the effective Hamiltonian in order to turn it self-adjoint in this frame, and the consequences for the cosmological scenario. One of the main problems with the Jordan frame is that the eq. \eqref{ej} is not separable in the coordinates $a$ and $\phi$. The avoidance of this inconvenience is one reason to use Einstein's frame. We can always change the coordinates of the solution \eqref{general solution} and it becomes a solution of \eqref{ej}.
However, it is not sufficient to ensure the equivalence between the solution in the two frames. It still remains to verify the conditions of self-adjointness of the Hamiltonian operator
\begin{eqnarray}
\label{Hamilt. 1}
\hat{H}_{J} = \frac{\omega}{12 (3+2\omega)} \biggr\{ \frac{1}{\phi} \left[\partial_a^2 + \frac{p}{a}\partial_a \right] + \frac{6}{\omega} \left[ \frac{1}{a} \partial_a\partial_\phi -\frac{1}{a^2}\left(\phi\partial_\phi^2 + q\partial_\phi \right) \right] \biggl\}
\end{eqnarray}
in Jordan frame, which will depend on the ordering factor \cite{moniz2}. In fact, \cite{pal} showed the equivalence of solutions for the two frames in an anisotropic model and acknowledge the importance of the ordering factors to ensure equivalence between both frames, but they did not analyse explicitly the reason behind it and the consequences of this restriction for the observables, which we intend to do here.

The first step is to determine for which measure the Hamiltonian \eqref{Hamilt. 1} is symmetric.
We suppose now that the inner product takes the form,
\begin{eqnarray}
(\xi,\Psi) = \int \xi^* \Psi a^r \phi^s da\,d\phi \,,
\end{eqnarray}
where the parameters $r$ and $s$ must be obtained imposing that $\hat{H}_{J}$ is symmetric, that is:
\begin{eqnarray}
(\xi,\hat{H}_{J} \Psi) = \int \xi^* (\hat{H}_{J} \Psi)a^r\phi^s da\,d\phi = \int (\hat{H}_{J} \xi^*) \Psi a^r\phi^s da\,d\phi = (\hat{H}_{J} \xi,\Psi) \,.
\end{eqnarray}
Performing explicitly the calculation, considering that the wavefunctions and their derivatives are zero at the boundaries, we obtain,
\begin{eqnarray}
(\xi,\hat{H}_{J} \Psi) &=& (\hat{H}_{J} \xi,\Psi)
\nonumber
\\
&+& \int \xi^*\,\Psi\biggr\{(r - p)(r - 1) + \frac{6}{\omega}s\biggr[(r - 1) - (s + 1 - q)\biggr]\biggl\}\,
\nonumber
\\
&\times& a^{r -2}\,\phi^{s - 1}\,da\,d\phi
\nonumber
\\
&+& \int (\partial_\phi \xi^*)\frac{6}{\omega}\biggr[ (r - 1) - 2(s + 1 - q)\biggl]\Psi\,a^{r - 2}\,\phi^s\,da\,d\phi
\nonumber
\\
&+& \int (\partial_a\xi^*)\biggr[2(r - p) + 6\frac{s}{\omega}\biggl]\Psi\,a^{r - 1}\,\phi^{s - 1}\,da\,d\phi \,.
\end{eqnarray}
Hence, the operator $\hat{H}_{J}$ is symmetric if the following conditions are satisfied:
\begin{eqnarray}
(r - p) + 3\frac{s}{\omega} &=& 0 \,;
\\
(r - 1) - 2[(s + 1) - q] &=& 0 \,;
\\
(r - p)(r - 1) + \frac{6}{\omega}s[((r- 1) - (s + 1) + q] &=& 0 \,.
\end{eqnarray}
This system of equation has a unique solution:
\begin{equation}
\label{r,s}
r = \frac{2p - 2(3-2q)}{(3+2\omega)} + (3 - 2q) \quad; \quad s= \frac{p-(3-2q)}{(3+2\omega)} \,.
\end{equation}
As in the case of Einstein frame, the measure depends on the ordering factors $p$ and $q$, and also on $\omega$.

We conclude, then, that in Jordan frame the Hamiltonian operator \eqref{Hamilt. 1} is symmetric with the measure $d\mu_{J} = a^{r} \phi^{s} \, da\,d\phi$, where $r$ and $s$ are given by \eqref{r,s}.
If we want to have square-integrable solution we must fix $q = 1$ \cite{moniz2}, and equation \eqref{r,s} becomes,
\begin{equation}
r = \frac{2(p+\omega) +1}{3 + 2\omega} \quad; \quad s= \frac{p-1}{3+ 2\omega} \,,
\end{equation}
where we have considered the transformation $da\,d\phi = \phi^{-\frac{1}{2}} db\,d\sigma$. However, it must be stressed that it is more convenient to work with the variables $(b,\sigma)$ instead of $(a,\phi)$. The measure $d\mu_{J}$ changes as
\begin{equation}
\label{medida Jordan -> Einstein}
a^{r} \phi^{s} \, da \, d\phi \quad \mapsto \quad b^{r} e^{-r\sigma/2}\,e^{s\sigma} \, e^{\sigma/2} db \,d\sigma = b^{r} e^{\sigma(2s-r+1)/2} \,db \, d\sigma = b^{r} db \, d\sigma \,.
\end{equation}
Notice that we suppress the primes in the expressions containing $\sigma$.
Therefore, we have
\begin{equation}
\label{measure Jordan}
d\mu_{J} = b^{r} \, db \, d\sigma \,,
\end{equation}
which is equivalent to $d\mu_{E}$ only for $r= \bar p = p = 1$.

The self-adjoint condition is related to the fact that we must have square-integrable solutions for an eigenvalue equation (see \cite{moniz2}). However, for an arbitrary $r$, the wavepacket $\Psi$ given by \eqref{wavepacket} is not square-itegrable. In fact:
\begin{eqnarray}
\nonumber
N &=& \int_{0}^{\infty} \int_{0}^{\infty} \Psi^{*} \, \Psi b^{r} db \, d\sigma
\\
&=& \int_{-\infty}^{+\infty}\int_0^{+\infty}e^{- 2k^2}\frac{b^{2\nu + 1 - \bar p +r}}{(4\alpha\alpha^*)^{\nu + 1}} \, e^{- \gamma \frac{b^2}{4\alpha\alpha^*}} \, db\, dk\,,
\end{eqnarray}
where we have already integrated over $\sigma$ and $k^{\prime}$. Changing the variables as in \eqref{variable y}, we obtain
\begin{equation}
\label{norm Jordan}
N = B^{\frac{\bar p - r}{2}} \int_{-\infty}^{+\infty}\int_0^{+\infty}e^{- 2k^2} \frac{y^{2\nu + 1 - \overline{p} + r}}{4^{\nu + 1}}e^{- \gamma \left(\frac{y}{2} \right)^{2}} \, dy \, dk\,.
\end{equation}
Recall that $B= \gamma^{2} + T^{2}$. Hence, the norm $N$ will be time-independent only if $\bar p = p = r = 1$. Consequently, the Hamiltonian in Jordan frame is only self-adjoint under this condition, which is the same condition to have the complete equivalence between the two frames.

We can also verify this equivalence by calculating  the inverse transformation of \eqref{medida Jordan -> Einstein}, that is, by writting $d\mu_{E}$ with the coordinates $(a, \phi)$. In this case, we have
\begin{equation}
\label{medida Einstein -> Jordan}
b^{\bar p} db \, d\sigma \quad \mapsto \quad a^{\bar p}\,\phi^{\bar p/2} \,\phi^{-1/2}\, da \, d\phi = a^{\bar p}\phi^{(\bar p-1)/2} \,da \, d\phi \,,
\end{equation}
which is equivalent to $d\mu_{J}$ only if $\overline{p}=p=1$, which means $r=1$ and $s=0$. Thus, the equivalence between Einstein and Jordan frames implies in the choice of a privileged factor ordering imposed in the quantization. We must have $p=1$ (implying $\overline{p}= 1$ in the Einstein frame) to obtain a self-adjoint operator in the Jordan frame as well.

Let us now calculate the expectation value of the scale factor $a$ in the Jordan frame. As commented above, it is better to use the coordinates $(b, \sigma)$. Then, $d\mu_{J}$ is given by \eqref{measure Jordan}. Letting $r$ arbitrary, we get
\begin{eqnarray}
\label{int-j}
\langle a \rangle = \langle \, b e^{-\frac{\sigma}{2}} \,\rangle = \frac{1}{N} \int_{-\infty}^{+\infty} \int_0^{\infty} \Psi^* \, b\,e^{-\frac{\sigma}{2}} \, \Psi \, b^{r} db \,d\sigma \,,
\end{eqnarray}
Substituting the wavepacket $\Psi$ \eqref{wavepacket}, the expression (\ref{int-j}) above takes the form,
\begin{eqnarray}
\langle a \rangle &=& \frac{1}{N} \int_{-\infty}^{+\infty} \int_{-\infty}^{+\infty} \int_{-\infty}^{+\infty} \int_0^{+\infty} e^{- k^2 - {k'}^2} \, e^{[i(k-k^{\prime}) - \frac{1}{2}]\sigma}
\nonumber
\\
&\times& \frac{b^{\nu +{\nu'}^* + 1 - \bar p + r}}{(2\alpha)^{\nu + 1}(2\alpha^*)^{{\nu'}^* + 1}} \,e^{- \gamma\frac{b^2}{2\alpha\alpha^*}} \, db \, d\sigma \, dk\,dk'\,,
\label{expected value a Jordan}
\end{eqnarray}
which is divergent due to the integration in $\sigma$. That is, $\langle a \rangle$ is divergent in both Einstein and Jordan frames. For the original field $\phi$, we have
\begin{equation}
\langle \phi \rangle = \langle e^{\sqrt{|\varpi|} \sigma} \rangle  = \frac{1}{N} \int_{-\infty}^{+\infty} \int_0^{\infty} \Psi^* \, e^{\sqrt{|\varpi|} \sigma} \, \Psi \, b^{r} db \,d\sigma \,,
\end{equation}
which also diverges because of the integral over $\sigma$. The quantities $\langle b \rangle$ and $\langle \sigma \rangle$ converges in both frames, nevertheless. In fact, for the expected value of $\sigma$, the integral over $k$ is still odd, even for an arbitrary $r$. Thus
we obtain
\begin{equation}
\langle \sigma \rangle = 0 \,,
\end{equation}
even in Jordan frame. However, as observed at the end of section \ref{section EF}, these results depend on the form (in fact, on the parity), of the superposition factor $A(k)$. On the other hand, in a similar way as we did in \eqref{expected value b}, and using the norm \eqref{norm Jordan} to obtain the expectation value for $b$, we obtain
\begin{equation}
\langle b \rangle  \propto \left( \sqrt{\gamma^2 + T^2} \right)^{1 + \bar p - r} \,,
\end{equation}
which reduces to
\begin{equation}
\langle b \rangle  \propto \sqrt{\gamma^2 + T^2} \,,
\end{equation}
when $\bar p = 1$ ($p = 1$).

\section{Conclusion}

We have analysed the self-adjoint character of the Hamiltonian obtained from the Brans-Dicke theory in mini-superspace, coupled with a radiative fluid, both in the Einstein and in the original Jordan frame, which are related by a conformal transformation. The radiative fluid is introduced through Schutz formalism, which allows us to obtain a Schr\"odinger-type equation with the fluid playing the r\^ole of time. In the Einstein frame, this problem has already been studied in Ref. \cite{moniz2}, and the self-adjoint properties of the resulting Hamiltonian operator have be classified in terms of the ordering parameter. It has been identified cases where the Hamiltonian operator is already self-adjoint, others where the Hamiltonian is not self-adjoint but admits a self-adjoint extension, and even a large class of cases where the Hamiltonian operator is neither self-adjoint, nor admits a self-adjoint extension.

In this paper, we have completed the analysis of Ref. \cite{moniz2}, determining the cosmological scenarios emerging from the quantum model in the Einstein frame.
It has been shown that a bouncing, non-singular Universe is a general prediction of the quantum model, with the expectation value of the scale factor depending very weakly on the self-adjoint properties of the Hamiltonian (essentially only some constants change for the different cases), and also on the wave packet construction. On the other hand, the scalar field changes considerably its functional form according to the different cases, and mainly according the superposition chosen to construct the wave packet.

The self-adjoint character of the Hamiltonian operator has been also analysed in the original Jordan frame. It has been found that the norm of the wavefunction is only time independent for a specific value of the ordering parameter. This value coincides with the one yielding an equivalence of both frames, the Einstein and Jordan frames. Moreover, in the Jordan frame the expectation values of the scale factor and of the scalar field converge only if the corresponding variables defined in the Einstein frame are used. This suggests that, at the quantum level, the Einstein frame is the physical one, which is a surprising result.

An immediate question concerns the generality of the results reported above. The Schr\"odinger-like equation is a partial differential equation, which can be solved using the separation of variable method, ensuring that all the solutions are separable \cite{Strauss}. The general solution can be constructed through arbitrary superpositions of the basis functions. In this paper, we used a class of these superpositions in order to construct the wave packets, and a specific choice among this class does not affect the convergence of the expectation value (specially for the scale factor), and we believe that any superposition which has the good properties at infinity will lead to the same results. Of course, the fact that we are restricted to mini-superspace models represent a clear limitation, and a more general set may change the conclusions made above. Also, the case $\epsilon > 0$ should be analyzed, even if the stability seems not to be assured for this case due to the non-positivity of the energy..
One possible way to investigate this is to add new degrees of freedom, considering for example anisotropic models. The self-adjoint properties of anisotropic models have been investigate in reference \cite{pal1,pal2,pal3}, and it would be interesting to consider anisotropies in the context of the Brans-Dicke theory. We hope to present such analysis in the future.

{\bf Acknowledgments:} We thank CNPq (Brazil) and FAPES (Brazil) for partial financial support.

\end{document}